\documentstyle [amsfonts,11pt]{article}

\def\ber#1#2{\begin{equation}\begin{array}{#1}\displaystyle{#2}}
\def\ber#1{\begin{equation}\begin{array}{#1}\displaystyle}
\def\bernn#1#2{$$\begin{array}{#1}\displaystyle{#2}}

\def\eer#1{\end{array}\label{#1}\end{equation}}
\def\eernn{\end{array}$$}
\def\r#1#2{\noindent\hbox{\hbox to 24 pt{\hfil[#1]~}%
\vtop{\hsize = 12.5 truecm\noindent#2}}\vskip 5 pt\vfil}
\def\chap#1#2#3{\noindent\hbox{\hbox to 1.5 truecm{\hfil#1}%
\hbox to 14 truecm{~#2\leaders\hbox to 0.5 em{\hfil.\hfil}\hfill#3}}\par}
\def\cchap#1#2#3#4{\noindent\hbox{\hbox to 1.5 truecm{\hfil#1}%
\hbox to 14 truecm{~#2\hfil}}\par
\noindent\hbox{\hskip 1.5 truecm%
\hbox to 14 truecm{~#3\leaders\hbox to 0.5 em{\hfil.\hfil}\hfill#4}}\par}
\def\bbt{\bibitem}
\def\be{\begin{equation}}
\def\en{\end{equation}}
\def\ber{\begin{eqnarray}}
\def\enr{\end{eqnarray}}
\def\nmb{ \nonumber\\}
\def\d{\partial}
\def\rbr{\rbrack}
\def\lbr{\lbrack}
\def\rbrc{\rbrace}
\def\lbrc{\lbrace}
\def\ov{\over }
\def\tld{\tilde}
\def\brv{\breve}

\def\MTR{Manin triple }
\def\MTRs{Manin triples }
\def\DLG{double Lie group }
\def\Tta{\Theta}
\def\sgm{\sigma}

\def\im{\imath}
\def\rh{\rho}

\begin{document}
\rightline{Landau Tmp/05/97.}
\rightline{May 1997}
\vskip 2 true cm
\centerline{\bf POISSON-LIE T-DUALITY AND
N=2 SUPERCONFORMAL WZNW MODELS}
\centerline{\bf ON COMPACT GROUPS.}
\vskip 2.5 true cm
\centerline{\bf S. E. Parkhomenko}
\centerline{Landau Institute for Theoretical Physics}
\centerline{142432 Chernogolovka,Russia}
\vskip 0.5 true cm
\centerline{spark@itp.ac.ru}
\vskip 1 true cm
\centerline{\bf Abstract}
\vskip 0.5 true cm
 The supersymmetric generalization of Poisson-Lie T-duality
in N=2 superconformal WZNW models on the compact groups is
considered. It is shown that the role of Drinfeld's doubles
play the complexifications of the corresponding compact groups.
These complex doubles are used to define the natural actions
of the isotropic subgroups forming the doubles on the group manifolds of
the N=2 superconformal WZNW models. The Poisson-Lie T-duality
in N=2 superconformal $U(2)$-WZNW model considered in details.
It is shown that this model admits Poisson-Lie symmetries with respect
to the isotropic subgroups forming Drinfeld's double $Gl(2,C)$.
Poisson-Lie T-duality transformation maps this model into itself but
acts nontrivialy on the space of classical solutions.
Supersymmetric generalization of Poisson-Lie T-duality in
N=2 superconformal WZNW models on the compact groups of higher
dimensions is proposed.
\smallskip
\vskip 10pt
\centerline{\bf Introduction.}
Target space duality in string theory has attracted a considerable
attention in recent years because it sheds some light on
the geometry and symmetries of string theory. The well known
example of T-duality is  mirror symmetry in the Calaby-Yao
manifolds compactifications of the superstring ~\cite{MS}.
Duality symmetry
was first described in the context of toroidal compactifications
~\cite{BGS}. For the simplest case of single compactified dimension
of radius $R$, the entire physics of interacting theory is left
unchanged under the replacement $R\to \alpha /R$ provided one also
transforms the dilaton field $\phi \to \phi -\ln{(R/\sqrt{\alpha})}$
~\cite{AlO}.
The T-duality symmetry
was later extended to the case of nonflat conformal backgrounds
with some abelian isometry (abelian T-duality) in ~\cite{Bu}.

 The basic idea of the notion of non-abelian duality
~\cite{OsQ,GR} is to consider a conformal field theory with
a non-abelian symmetry group. The non-abelian duality did miss a lot
of features characteristic to the abelian duality. For example
the non-abelian T-duality transformation of the isometric
$\sigma$- model on a group manifold $G$ gives non-isometric
$\sigma$- model on its Lie algebra ~\cite{OsQ,FrJ}.
As a result, it was not known how to perform the inverse
duality transformation to get back to the original model.

 A solution of this problem was proposed recently in ~\cite{KlimS1},
where it was argued that the two theories are dual
to each other from the point of view of the so called Poisson-Lie (PL)
T-duality. The main idea of their approach is to replace
the requirement of isometry by a weaker condition
which is the Poisson-Lie symmetry of the theory. This generalized
duality is associated with two groups forming a Drinfeld double
~\cite{Drinf1} and the duality transformation exchanges their roles.
This approach has recieved futher developments in the series
of works ~\cite{KlimS2}, ~\cite{KlSWZ1}, ~\cite{KlSWZ2}, ~\cite{KlSM},
~\cite{TyU}.

 In order to apply PL T-duality in superstring
theory one needs to have the dual pairs of conformal
and superconformal $\sigma$-models.
The simple example of dual pair of conformal $\sigma$-models
associated with the $O(2,2)$ Drinfeld double was presented
in work ~\cite{AlKT}. Then, it was shown in ~\cite{KlSWZ1},
~\cite{KlSWZ2} that WZNW models on the compact groups are the
natural examples of PL dualizable $\sigma$-models.

 The supersymmetric generalization
of PL T-duality was considered in~\cite{Sfet,P}.
In particular, due to the close relation between
N=2 superconformal WZNW models and Drinfeld's double
(Manin triple) structures on the corresponding group manifolds
(Lie algebras) ~\cite{QFR3,QFR}, it was shown in the paper
~\cite{P} that N=2 superconformal WZNW models admit very natural
PL symmetry and PL T-dual $\sigma$-models for N=2 WZNW models
associated with real Drinfeld's doubles was constructed.

 In the present note we consider PL T-duality
in N=2 superconformal WZNW models on the compact groups.
These models correspond to the complex Manin triples endowed
with hermitian conjugation which conjugates isotropic
subalgebras of the Manin triple.

 After a brief review of the classical N=1 superconformal
WZNW models in the section 1, we consider in the
section 2, the Manin triple construction of N=2 superconformal
WZNW models on the compact groups and define the action of
isotropic subgroups from the correspondig Drinfeld's doubles
on the space of fields. In the section 3 we consider
N=2 superconformal $U(2)$-WZNW model. We shall show by the direct
calculations that this model admits Poisson-Lie symmetries with
respect to the action of the isotropic subgroups forming Drinfeld's
double. Using these symmetries we show that Poisson-Lie T-dual
$\sigma$-model is N=2 $U(2)$-WZNW model itself. We conclude with
some proposals about extensions of our results for N=2 superconformal
WZNW models on the other compact groups and Kazama-Suzuki models.

\vskip 10pt
\centerline{\bf1. The classical N=1 superconformal WZNW model.}

 In this section we briefly review a superconformal
WZNW (SWZNW) models using superfield formalism ~\cite{swzw}.

 We parametrize super world-sheet introducing the light cone
coordinates
$x_{\pm}$, and grassman coordinates $\Tta_{\pm}$.
The generators of supersymmetry and covariant
derivatives are
\be
Q_{\mp}= {\d \ov \d\Tta_{\pm}}+\im \Tta_{\pm}\d_{\mp},\
D_{\mp}= {\d \ov \d\Tta_{\pm}}-\im \Tta_{\pm}\d_{\mp}.
\label{1}
\en
They satisfy the relations
\be
\lbrc D_{\pm},D_{\pm}\rbrc= -\lbrc Q_{\pm},Q_{\pm}\rbrc= -\im 2\d_{\pm},\
\lbrc D_{\pm},D_{\mp}\rbrc= \lbrc Q_{\pm},Q_{\mp}\rbrc=
\lbrc Q,D\rbrc= 0,
\label{2}
\en
where the brackets $\lbrc,\rbrc$ denote the anticommutator.
The superfield of N=1 supersymmetric WZNW model
\be
G= g+ \im \Tta_{-}\psi_{+}+ \im \Tta_{+}\psi_{-}+
   \im \Tta_{-}\Tta_{+}F  \label{3}
\en
takes values in a Lie group ${\bf G}$.
We will assume that its Lie algebra ${\bf g}$
is endowed with ad-invariant nondegenerate inner
product $<,>$.

The inverse group element $G^{-1}$ is defined from the relation
\be
 G^{-1}G=1 \label{4}
\en
and has the decomposition
\be
 G^{-1}= g^{-1}- \im \Tta_{-}g^{-1}\psi_{+}g^{-1}-
         \im \Tta_{+}g^{-1}\psi_{-}g^{-1}-
         \im \Tta_{-}\Tta_{+}g^{-1}(F+\psi_{-}g^{-1}\psi_{+}-
         \psi_{+}g^{-1}\psi_{-})g^{-1} \label{5}
\en

 The action of N=1 SWZNW model is given by
\ber
S_{swz}= \int d^{2}x d^{2} \Tta(<R_{+},R_{-}>)   \nmb
         -\int d^{2}x d^{2}\Tta dt
          <G^{-1}\frac{\d G}{\d t},\lbrc R_{-},R_{+}\rbrc>, \label{7}
\enr
where
\be
 R_{\pm}= G^{-1}D_{\pm}G.  \label{8}
\en
The classical equations of motion can be obtained by making a variation
of (\ref{7}):
\be
\delta S_{swz}= \int d^{2}x d^{2} \Tta
<G^{-1}\delta G,D_{-}R_{+}-D_{+}R_{-}-\lbrc R_{-},R_{+}\rbrc>
\label{9}
\en
Taking into account kinematic relation
\be
D_{+}R_{-}+D_{-}R_{+}= -\lbrc R_{+},R_{-}\rbrc \label{rel}
\en
we obtain
\be
 D_{-}R_{+}=0
\label{10}
\en

The action (\ref{7}) is invariant under the super-Kac-Moody
and N=1 superconformal
transformations ~\cite{swzw}.

 In the following we will use supersymmetric version of
Polyakov-Wiegman formula ~\cite{PW}
\be
S_{swz}[GH]= S_{swz}[G]+ S_{swz}[H]+ \int d^{2}x d^{2}\Tta
             <G^{-1}D_{+}G,D_{-}HH^{-1}>.  \label{11}
\en
It can be proved as in the non supersymmetric case.

\vskip 10pt
\centerline{\bf2. N=2 superconformal WZNW models on compact groups.}

 In works ~\cite{QFR2,QFR3,QFR} supersymmetric WZNW models which admit
extended supersymmetry were studied and correspondence between extended
supersymmetric WZNW models and finite-dimensional Manin triples was
established in ~\cite{QFR3,QFR}.
By the definition ~\cite{Drinf1},
a \MTR $({\bf g},{\bf g_{+}},{\bf g_{-}})$
consists
of a Lie algebra ${\bf g}$, with nondegenerate invariant inner product
$<,>$ and isotropic Lie subalgebras ${\bf g_{\pm}}$ such that
${\bf g}={\bf g_{+}}\oplus {\bf g_{-}}$ as a vector space.

 The corresponding Sugawara construction of N=2 Virasoro superalgebra
generators was given in ~\cite{QFR3,QFR,QFR2,GETZ}.

 To make a connection between Manin triple construction of
~\cite{QFR3,QFR} and approach of ~\cite{QFR2} based on complex
structures on Lie algebras the following comment is relevant.

 Let ${\bf g}$ be a real Lie algebra and $J$ be a complex  structure
on the vector space ${\bf g}$. $J$ is referred to as the complex
structure
on the Lie algebra ${\bf g}$ if $J$ satisfies the equation
\be
\lbr Jx,Jy \rbr-J\lbr Jx,y \rbr-J\lbr x,Jy \rbr=\lbr x,y \rbr \label{12}
\en
for any elements $x, y$ from ${\bf g}$.
It is clear that corresponding Lie group is
complex manifold with left (or right) invariant complex structure.
In the following we shall denote a real Lie group
and real Lie algebra
with a complex structure satisfying (\ref{12}) as
the pairs $({\bf G}, J)$ and $({\bf g}, J)$ correspondingly.

 Suppose the existence of a nondegenerate invariant inner product $<,>$ on
${\bf g}$ so that the complex structure $J$ is skew-symmetric with
respect to $<,>$. In this case it is not difficult to establish the
correspondence between complex \MTRs and complex structures on
Lie algebras. Namely, for each complex \MTR
$({\bf g},{\bf g_{+}},{\bf g_{-}})$
exists a canonic complex structure on the Lie algebra ${\bf g}$ such
that subalgebras ${\bf g_{\pm}}$ are its $\pm \im$ ei\-gen\-spa\-ces.
On the other hand, for each real Lie algebra ${\bf g}$
with nondegenerate invariant inner product and
skew-symmetric complex structure $J$ on this algebra one can
consider the complexification ${\bf g^{\Bbb C}}$ of ${\bf g}$. Let
${\bf g_{\pm}}$ be $\pm \imath$ eigenspaces of $J$ in the
algebra ${\bf g^{\Bbb C}}$ then $({\bf g^{\Bbb C}},{\bf g_{+}},{\bf g_{-}})$
is a complex \MTR.
 Moreover it can be prooved ~\cite{QFR3} that there exists one-to-one
correspondence between the complex Manin triple endowed with antilinear
involution which conjugates isotropic subalgebras
$\tau: {\bf g_{\pm}}\to
{\bf g_{\mp}}$ and the real Lie
algebra endowed with $ad$-invariant nondegenerate inner product $<,>$
and the complex structure $J$ which is skew-symmetric with respect
to $<,>$. Therefore we can use this conjugation to extract a real
form from the complex Manin triple.

 If the complex structure on the Lie algebra is fixed then it defines the
second supersymmetry transformation ~\cite{QFR2}.

In this paper we concentrate on  N=2 SWZNW models
on the compact groups, 
that is we shall consider complex Manin triples
such that the corresponding antilinaer involutions will
coincide with the hermitian conjugations. Hence it will be implied
in the following that ${\bf G}$ is a subgroup in the group of
unitary matrices and the matrix elements of
the superfield $G$ satisfy the relations:
\be
\bar{g}^{mn}=(g^{-1})^{nm},\
\bar{\psi}^{mn}_{\pm}= (\psi^{-1})^{nm}_{\pm},\
\bar{F}^{mn}= (F^{-1})^{nm}, \label{6.u}
\en
where we have used the following notations
\be
\psi^{-1}_{\pm}= -g^{-1}\psi_{\pm}g^{-1},\
F^{-1}= -g^{-1}(F+\psi_{-}g^{-1}\psi_{+}-
         \psi_{+}g^{-1}\psi_{-})g^{-1}. \label{6.not}
\en

 Now we have to consider some geometric properties of the N=2 SWZNW
models closely related with the existence of a complex structure
$J$ on the corresponding groups.

 Let's fix some compact Lie group with a complex structure $({\bf G}, J)$
and consider its Lie algebra with the complex structure $({\bf g}, J)$.
The complexification ${\bf g^{\Bbb C}}$ of ${\bf g}$ has the Manin triple
structure $({\bf g^{\Bbb C}},{\bf g_{+}},{\bf g_{-}})$. The Lie group version
of this triple ~\cite{SemTian,LuW,AlMal} is the
\DLG $({\bf G^{\Bbb C}},{\bf G_{+}},{\bf G_{-}})$ ~\cite{LuW}, where the
exponential groups ${\bf G_{\pm}}$ correspond to the Lie algebras
${\bf g_{\pm}}$. The real Lie group ${\bf G}$ is extracted
from its complexification with help of hermitian conjugation $\tau$
\be
{\bf G}= \lbrc g\in {\bf G^{\Bbb C}}|\tau (g)=g^{-1}\rbrc       \label{rf}
\en
 Each element $g\in {\bf G^{\Bbb C}}$ from the
vicinity of the unit element from ${\bf G^{\Bbb C}}$
admits two decompositions
\be
g= g_{+}g^{-1}_{-}= {\tld g}_{-}{\tld g}^{-1}_{+},  \label{13}
\en
where ${\tld g}_{\pm}$ are dressing transformed
elements of $g_{\pm}$ ~\cite{LuW}:
\be
{\tld g}_{\pm}=(g^{-1}_{\pm})^{g_{\mp}}         \label{13not}
\en
Taking into account (\ref{rf}) and (\ref{13}) we conclude that the
element $g$ (from the vicinity of unit element from ${\bf G}^{\Bbb C}$)
belongs to ${\bf G}$ iff
\be
\tau (g_{\pm})= {\tld g}^{-1}_{\mp}      \label{13u}
\en
These equations mean that we can parametrize the elements from ${\bf G}$
by the elements from the complex group ${\bf G}_{+}$ (or ${\bf G}_{-}$)
e.i. we can introduce complex coordinates (they are just matrix elements
of $g_{+}$ (or $g_{-}$)) in the vicinity of unit element from ${\bf G}$.
To do it one needs to solve with respect to $g_{-}$ the following equation:
\be
\tau (g_{-})= (g_{+})^{g^{-1}_{-}}              \label{13c+}
\en
(to introduce ${\bf G_{-}}$-coordinates on ${\bf G}$ one needs to solve
with respect to $g_{+}$ the equation
\be
\tau (g_{+})= (g_{-})^{g^{-1}_{+}} ).            \label{13c-}
\en
For the N=2 SWZNW model on the group ${\bf G}$ we obtain from
(\ref{13}) the decompositions for the superfield (\ref{4}) (which takes
values in the vicinity of unit element from ${\bf G}$)
\be
G(x_{+},x_{-})= G_{+}(x_{+},x_{-})G^{-1}_{-}(x_{+},x_{-})=
                {\tld G}_{-}(x_{+},x_{-}){\tld G}^{-1}_{+}(x_{+},x_{-})
\label{14}
\en
Due to (\ref{14}), (\ref{11}) and the definition of Manin triple we
can rewrite the action (\ref{7}) for this superfield
in the following manifestly real form
\be
S_{swz}=-{1\ov 2}\int d^{2}x d^{2}\Tta (<\rh^{+}_{+}, \rh^{-}_{-}>+
          <{\tld \rh}^{-}_{+}, {\tld \rh}^{+}_{-}>)  \label{15}
\en
where the superfields
\be
\rh^{\pm}= G^{-1}_{\pm}DG_{\pm}, \
{\tld \rh}^{\pm}= {\tld G}^{-1}_{\pm}D{\tld G}_{\pm}
\label{16}
\en
correspond to the left invariant 1-forms on the groups
${\bf G_{\pm}}$ and ${\tld \rh}^{\pm}$, $\rh^{-}$ are
expressed in terms of matrix elements of $G_{+}$ (and its complex
conjugated) with help of the formulas
(\ref{13}, \ref{13u}, \ref{13c+}).

 To generalize (\ref{13}), (\ref{13u}) we have to consider the
set $W$ of classes ${\bf G_{+}}\backslash {\bf G^{\Bbb C}}/ {\bf G_{-}}$
and pick up a representative $w$ for each class $[w]\in W$
(which we shall assume in the following to be discret set).
It gives us the following stratification
of ${\bf G^{\Bbb C}}$ ~\cite{AlMal}:
\be
{\bf G}^{\Bbb C}= \bigcup_{[w]\in W} {\bf G_{+}}w{\bf G_{-}}=
         \bigcup_{[w]\in W} {\bf G_{w}}  \label{17+}
\en
There is the second stratification:
\be
{\bf G^{\Bbb C}}= \bigcup_{[w]\in W} {\bf G_{-}}w{\bf G_{+}}=
         \bigcup_{[w]\in W} {\bf G^{w}}  \label{17-}
\en
We shall assume that the representatives $w$
have picked up to satisfy the unitarity condition:
\be
\tau (w)=w^{-1}                        \label{17w}
\en
This allows us to generalize (\ref{13}), (\ref{13u}) as follows
\be
g= g_{+}wg^{-1}_{-}= {\tld g}_{-}w{\tld g}^{-1}_{+},    \label{18}
\en
where $g_{\pm}$ related with ${\tld g}_{\pm}$ by the formula
(\ref{13u}). The corresponding generalization of (\ref{14}) allows
us to write down the following generalization of (\ref{15})
\be
S_{swz}=-{1\ov 2}\int d^{2}x d^{2}\Tta (<\rh^{+}_{+}, w\rh^{-}_{-}w^{-1}>+
          <{\tld \rh}^{-}_{+}, w{\tld \rh}^{+}_{-}w^{-1}>)  \label{19}
\en
It is clear that the formula (\ref{19}) is correct inside the world
sheet domain where the superfields take values in the class ${\bf G_{w}}$.
On the boundaries of these domains, where the jumps from one class to
another is appeared
some additional terms should be added, but for our purposes it will
suffice to consider the mappings when the whole world-sheet maps
into the one of the classes from $W$.

 The formulas (\ref{13c+}) ((\ref{13c-})), (\ref{18}), (\ref{19})
mean that there is a natural action of the complex group ${\bf G_{+}}$
(${\bf G_{-}}$) on ${\bf G}$, and the set $W$ parametrizes
${\bf G_{+}}$-orbits (${\bf G_{-}}$-orbits)
${\bf G}\cap {\bf G_{w}}$. It is easy to see also that
the group ${\bf G^{\Bbb C}}$ plays the role of Drinfeld's double for
the compact group with the complex structure $({\bf G},J)$.
Therefore it would be well to represent a supersymmetric generalization
of an off shell PL T-duality formulation using this double. Although the
off shell formulation is known in bosonic case ~\cite{KlimS2},
~\cite{KlSWZ1} its supersymmetric generalization is not straightforward
and is still an open problem. Consequently we shall work with the
formulas (\ref{13c+}, \ref{14}, \ref{17w}, \ref{18}, \ref{19}) and
propose an on shell supersymmetric version of PL T-duality in N=2 SWZNW
models on the compact groups considering the simplest
example- N=2 $U(2)$-SWZNW model.

\vskip 10pt
\centerline{\bf 3. Poisson-Lie T-duality in N=2 $U(2)$-SWZNW model.}

\vskip 5pt
{\it 3.1. The complex structure, Manin triple and \DLG for $U(2)$}\

 Let's fix the standard basis in the Lie algebra ${\bf g}$ of the group
$U(2)$:
\be
\sgm _{0}=\left(\begin{array}{cc}
                 \im &0\\
                 0&\im
                 \end{array}\right),\
\sgm _{1}=\left(\begin{array}{cc}
                 0&1\\
                 -1&0
                 \end{array}\right),\
\sgm _{2}=\left(\begin{array}{cc}
                 0&\im \\
                 \im &0
                 \end{array}\right),\
\sgm _{3}=\left(\begin{array}{cc}
                 \im &0\\
                 0&-\im
                 \end{array}\right)
\label{20}
\en
We introduce the complex structure $J$ on ${\bf g}$ by the formulas
\be
J\sgm _{0}=\sgm _{3},\
J\sgm _{1}=\sgm _{2}
\label{21}
\en
and define its ${\pm}\im $-basics eigenvectors:
\be
e^{0}={1\ov 2}(1-\im J)\sgm _{0},\
e^{1}={1\ov 2}(1-\im J)\sgm _{1}
\label{22-}
\en
\be
e_{0}={1\ov 2}(1+\im J)\sgm _{0},\
e_{1}={1\ov 2}(1+\im J)\sgm _{0}
\label{22+}
\en
Let's define the following subspaces in the complexification ${\bf g^{\Bbb C}}$
of the Lie algebra ${\bf g}$
\be
{\bf g_{-}}={\Bbb C}e^{0}+{\Bbb C}e^{1},\
{\bf g_{+}}={\Bbb C}e_{0}+{\Bbb C}e_{1}
\label{23}
\en
Thus ${\bf g^{\Bbb C}}={\bf g_{-}}\oplus {\bf g_{+}}$ as a vector space
and ${\bf g_{\pm}}$ are maximaly isotropic subalgebras with respect
to the inner product $<,>$ on ${\bf g^{\Bbb C}}$ defined by
\be
<x,y>= Tr(xy), \label{24}
\en
where $x, y\in {\bf g^{\Bbb C}}$. Therefore we have obtained the Manin triple
\be
({\bf g^{\Bbb C}}, {\bf g_{-}}, {\bf g_{+}})  \label{25}
\en
so that the vectors (\ref{22-}, \ref{22+}) constitute the orthonormal
basis in ${\bf g^{\Bbb C}}$.

 We parametrize the elements from the exponential subgroups
${\bf G_{\pm}}=Exp({\bf g}_{\pm})$ by the matrices
\be
g_{+}=\left(\begin{array}{cc}
                 a_{+}&0\\
                 -a_{+}b_{+}&a_{+}^{-1}
                 \end{array}\right)a_{+}^{-\im}, \
g_{-}=\left(\begin{array}{cc}
                 a_{-}&a_{-}^{-1}b_{-}\\
                 0&a_{-}^{-1}
                 \end{array}\right)a_{-}^{\im},
\label{26}
\en
where
\be
a_{\pm}=\exp({\mp}z_{\pm}/2)   \label{27}
\en
(in the following we shall use the same notations: $a_{\pm}, a_{\pm},
z_{\pm}$ for the matrix elements of the superfields $G_{\pm}$).
We denote by $({\bf G}^{\Bbb C},{\bf G_{+}},{\bf G_{-}})$ the \DLG,
where
\ber
{\bf G^{\Bbb C}}={\bf G^{\Bbb C}_{1}}\cup {\bf G^{\Bbb C}_{s}}\sim Gl(2,C), \nmb
{\bf G^{\Bbb C}_{1}}={\bf G_{+}}{\bf G_{-}},\
{\bf G^{\Bbb C}_{s}}={\bf G_{+}}\left(\begin{array}{cc}
                                      0&1\\
                                     -1&0
                                      \end{array}\right){\bf G_{-}}.
\label{28}
\enr
 The solution of the equations (\ref{13c+}) are given by
\be
z_{-}=-\bar{z}_{+}+\ln(1+\mid b_{+}\mid ^{2}),\
b_{-}=-\exp(z_{+}-\bar{z}_{+})\bar{b}_{+}
\label{29}
\en
and the solution of the equations (\ref{13}) are given by
\ber
\tld{z}_{+}=-z_{+}+\ln(1+\mid b_{+}\mid ^{2}),\
\tld{z}_{-}=\bar{z}_{+} \nmb
\tld{b}_{+}=-\exp(\bar{z}_{+}-z_{+})b_{+},\
\tld{b}_{-}=\bar{b}_{+}
\label{30}
\enr
Using these formulas we can represent the action (\ref{15}) in the
components
\be
S_{swz}=-{1\ov 2}\int d^{2}x d^{2}\Tta (E_{ij}\rh^{i}_{+}\rh^{j}_{-}+ \
E_{i\bar{j}}(\rh^{i}_{+}\bar{\rh^{j}_{-}}-\rh^{i}_{-}\bar{\rh^{j}_{+}})+ \
E_{\bar{i}\bar{j}}\bar{\rh^{i}_{+}}\bar{\rh^{j}_{-}}),
\label{31}
\en
where $i,j= 0,1$,
\ber
E_{01}=-E_{10}=-(1+\mid b_{+}\mid^{2})^{-1}a_{+}^{-2}\bar{b}_{+},\
E_{\bar{i}\bar{j}}=\bar{E}_{ij}       \nmb
E_{0\bar{0}}=1,\
E_{0\bar{1}}=-(1+\mid b_{+}\mid^{2})^{-1}\bar{a}_{+}^{-2}b_{+}   \nmb
E_{1\bar{0}}=-(1+\mid b_{+}\mid^{2})^{-1}a_{+}^{-2}\bar{b}_{+},\
E_{1\bar{1}}=(1+\mid b_{+}\mid^{2})^{-1}a_{+}^{-2}\bar{a}_{+}^{-2},
\label{32}
\enr
other elements of the bilinear form $E$ are zeroes and
$\rh^{i}$ are given by
\be
\rh^{0}=Dz_{+},\
\rh^{1}=a^{2}_{+}Db_{+}
\label{33}
\en
(the fields $\rh^{i}$ correspond to the components of the left
invariant 1-form $g^{-1}_{+}dg_{+}$ on the group ${\bf G_{+}}$).
For the mappings into the class ${\bf G^{\Bbb C}_{s}}$ the action (\ref{19})
is identicaly zero because
\be
<\rh^{+}_{+}, w\rh^{-}_{-}w^{-1}>= \
<{\tld \rh}^{-}_{+}, w{\tld \rh}^{+}_{-}w^{-1}>=0.
\label{34}
\en
\vskip 5pt
{\it 3.2. Poisson-Lie symmetry conditions.}\

 In view of the formulas (\ref{31}-\ref{34}) we can consider
$U(2)$-SWZNW model as a $\sgm$-model on the complex Lie group
${\bf G_{+}}$ and find the equations of motion making a variation
of the action (\ref{31}) under the right action of this group
on itself.

 The right translations on the group ${\bf G_{+}}$ are generated
by the vector fields
\be
S_{0}=\frac{\d}{\d z_{+}},\
S_{1}=\exp(z_{+})\frac{\d}{\d b_{+}}
\label{35}
\en
and its complex conjugated $\bar{S}_{i}, i=0,1$.
Note that the second vector field from (\ref{35}) coincides
with the classical screening current in the Wakimoto representations
of $\hat{sl}(2)$ ~\cite{Wak}. We would like to stress the general
feature of this observation: the screening currents in the Wakimoto
representations are given by the right action of the maximal nilpotent
subgroup ${\bf N_{+}}$ on the big cell of the corresponding flag
manifold ${\bf H/B_{-}}$, where ${\bf B_{-}}$ is the Borelian
subgroup of the group ${\bf H}$ , (${\bf N_{+}}\in {\bf B_{+}}$)
~\cite{scr}, consequently the natural action of the group ${\bf G_{+}}$
in any N=2 SWZNW model on the group ${\bf G}$ associated
with the Bruhat decomposition is generated by
the screening currents.

 Making a variation of (\ref{31}) under the vector field
$Z=Z^{i}S_{i}+Z^{\bar{i}}\bar{S}_{i}$ we obtain on the extremals
\ber
D_{+}(A_{-})_{i}+D_{-}(A_{+})_{i}-L_{S_{i}}E=0 \nmb
D_{+}(A_{-})_{\bar{i}}+D_{+}(A_{+})_{\bar{i}}-L_{\bar{S}_{i}}E=0,
\label{36}
\enr
where $L_{S_{i}}, L_{\bar{S}_{i}}$ mean the Lie derivatives along
the vector fields $S_{i}, \bar{S}_{i}$
and the Noether currents $A_{i}, A_{\bar{i}}$ are given by
\ber
(A_{-})_{i}=E_{i\bar{j}}\bar{\rh}^{j}_{-}+E_{ij}\rh^{j}_{-}, \nmb
(A_{+})_{i}=-E_{i\bar{j}}\bar{\rh}^{j}_{+}-E_{ji}\rh^{j}_{+}, \nmb
(\bar{A}_{\pm})_{i}=(A_{\pm})_{\bar{i}}
\label{37}
\enr
By the direct calculation we can check that the following PL symmetry
conditions are satisfied on the extremals
\ber
L_{S_{i}}E= f^{nm}_{i}(A_{+})_{n}(A_{-})_{m} \nmb
L_{\bar{S}_{i}}E= \bar{f}^{nm}_{i}(A_{+})_{\bar{n}}(A_{-})_{\bar{m}},
\label{38}
\enr
where $f^{nm}_{i}$ are the structure constants of the Lie algebra
${\bf g_{-}}$. By demanding the closure of (\ref{38}):
$\lbr L_{S_{i}},L_{S_{j}}\rbr=f_{ij}^{k}L_{S_{k}}$, we obtain
the following consistency condition
\be
f^{n}_{ij}f^{km}_{n}=f^{nm}_{j}f^{k}_{in}-f^{nk}_{j}f^{m}_{in} \
                     -f^{nm}_{i}f^{k}_{jn}+f^{nk}_{i}f^{m}_{jn}
\label{39}
\en
which is satisfyed due to the Jacoby identity in the Lie algebra
${\bf g^{\Bbb C}}$.

 As it is easy to see from (\ref{36}) the eq. (\ref{38}) are equivalent
to the zero curvature equations for the $F_{+-}$-component of the super
stress tensor $F_{MN}$
\ber
(F_{+-})_{i}\equiv D_{+}(A_{-})_{i}+D_{-}(A_{+})_{i}-
              f^{nm}_{i}(A_{+})_{n}(A_{-})_{m}=0  \nmb
(F_{+-})_{\bar{i}}\equiv D_{+}(A_{-})_{\bar{i}}+D_{-}(A_{+})_{\bar{i}}-
              \bar{f}^{nm}_{i}(A_{+})_{\bar{n}}(A_{-})_{\bar{m}}=0
\label{40}
\enr
Using the standard arguments of the super Lax construction ~\cite{EvHol}
one can show that from (\ref{40}) it follows that the connection is flat
\be
F_{MN}=0,\ M, N= (+, -, +, -).  \label{41}
\en
 The equations (\ref{40}) are the supersymmetric
generalization of Poisson-Lie symmetry conditions from the work
~\cite{KlimS1}. Indeed, the Noether currents $A_{i}, A_{\bar{i}}$
are generators of ${\bf g_{+}}$- action, while the
structure constants in (\ref{40}) correspond to Lie algebra
${\bf g_{-}}$ which is Drinfeld's dual to ${\bf g_{+}}$
~\cite{Drinf2}.
\vskip 5pt
{\it 3.3. Poisson-Lie T-dual $\sgm $-model.}\

 The PL-dual to $U(2)$-SWZNW $\sgm$-model should obey the conditions
as (\ref{38}) but with the roles of the Lie algebras ${\bf g_{\pm}}$
interchanged ~\cite{KlimS1}.

 To construct this $\sgm$-model we associate (due to (\ref{41}))
to each solution
$G_{+}(x_{+}, x_{-}, \Tta_{+}, \Tta_{-})$,
the map $V_{-}(x_{+}, x_{-}, \Tta_{+}, \Tta_{-})$ from the super
world-sheet into the group ${\bf G_{-}}$ such that
\be
(A_{\pm})_{i}e^{i}=D_{\pm}V_{-}V^{-1}_{-}
\label{42}
\en
From the other hand, using (\ref{29}, \ref{32}) one can obtain
\ber
(A_{-})_{i}e^{i}=-\rh^{-}_{-} \nmb
(A_{+})_{i}e^{i}\neq \pm \rh^{-}_{+}
\label{43}
\enr
Therefore one can represent $V_{-}$ as the product
\be
V_{-}=G^{-1}_{-}H^{-1}_{-}
\label{43p}
\en
, where $G_{-}$ is determined
from (\ref{14}) and $H_{-}$ satisfy the equation
\be
D_{-}H_{-}=0 \label{44}
\en

 Now we build the following surface in the Drinfeld's
double ${\bf G^{\Bbb C}}$:
\ber
F(x_{+}, x_{-}, \Tta_{+}, \Tta_{-})=
G_{+}(x_{+}, x_{-}, \Tta_{+}, \Tta_{-})
V_{-}(x_{+}, x_{-}, \Tta_{+}, \Tta_{-})\nmb
=G(x_{+}, x_{-}, \Tta_{+}, \Tta_{-})H_{-}(x_{+}, x_{-}, \Tta_{+}, \Tta_{-}),
\label{46}
\enr
where $G_{+}(x_{+}, x_{-}, \Tta_{+}, \Tta_{-}),
G(x_{+}, x_{-}, \Tta_{+}, \Tta_{-})$ are the classical solutions
of $U(2)$-SWZNW model,
$V_{-}(x_{+}, x_{-}, \Tta_{+}, \Tta_{-})$ satisfy (\ref{42})
and we have used (\ref{43p}).
The solution of the dual $\sgm $-model is given by "dual"
parametrization of the surface (\ref{46}) ~\cite{KlimS1}
\ber
F(x_{+}, x_{-}, \Tta_{+}, \Tta_{-})=
\brv{G}_{-}(x_{+}, x_{-}, \Tta_{+}, \Tta_{-})
\brv{V}_{+}(x_{+}, x_{-}, \Tta_{+}, \Tta_{-}) \nmb
=\brv{G}(x_{+}, x_{-}, \Tta_{+}, \Tta_{-})
\brv{H}_{+}(x_{+}, x_{-}, \Tta_{+}, \Tta_{-}),
\label{46d}
\enr
where $\brv{G}(x_{+}, x_{-}, \Tta_{+}, \Tta_{-})\in {\bf G}$
and $\brv{H}_{+}(x_{+}, x_{-}, \Tta_{+}, \Tta_{-})\in {\bf G_{+}}$.
In the dual $\sgm$-model Drinfeld's dual group to the group
${\bf G_{+}}$ should acts, i.e. it should be a $\sgm$-model on the
orbits of the group ${\bf G_{-}}$.
Because the hermitian conjugation $\tau$ conjugates
subgroups ${\bf G_{\pm}}$ and the action (\ref{19})
is real we conclude that the dual $\sgm$-model should
coincide with the initial one, thus N=2 $U(2)$-SWZNW model
is PL self-dual.


\vskip 10pt
\centerline{\bf5. Conclusions.}

 We have shown that the classical N=2 $U(2)$-SWZNW model possess
very natural PL symmetry with respect to the
isotropic subgroups ${\bf G_{\pm}}$ forming the Drinfeld's double
$Gl(2,C)$. The infinitesimal action of the group ${\bf G_{+}}$
(${\bf G_{-}}$) are given by the classical screening currents
in the Wakimoto representations of the $\hat{u}(2)$-current algebra.
PL T-dual $\sgm$-model to N=2 $U(2)$-SWZNW model is
N=2 $U(2)$-SWZNW itself but the PL T-duality transformation acts
nontrivialy on the classical solutions.

 We have shown also that the role of Drinfeld's double in N=2
SWZNW model on an arbitriary compact group ${\bf G}$ plays its
complexification ${\bf G^{\Bbb C}}$ endowed with the hermitian conjugation
which conjugates isotropic subgroups.
There is the natural action of isotropic subgroups ${\bf G_{\pm}}\in
{\bf G^{\Bbb C}}$ on ${\bf G}$ generated by the corresponding screening
currents. Having established PL symmetry in the simplest
case of N=2 $U(2)$-SWZNW model it would appear reasonable that
{\it N=2 SWZNW models on the compact groups of higher dimensions
admit PL symmetries with respect to the actions
of isotropic subgroups from the corresponding Drinfeld's doubles
such that these models are PL T-selfdual.}\ We will publish the
proof of this proposal in the nearst future ~\cite{pr}.
We expect the PL T-duality exists
also in N=2 superconformal Kazama-Suzuki models ~\cite{KaSu} since
these models can be represented as the cosets
(N=2 ${\bf G}$-SWZNW model)/(N=2 ${\bf H}$-SWZNW model),
where ${\bf H}$ is a subgroup from ${\bf G}$ ~\cite{HulS}.

 It is an intresting question what is the quantum picture of the
PL T-duality in N=2 SWZNW models. Because the Poisson-Lie groups
is nothing but a classical limit of the quantum groups ~\cite{Drinf1}
there appears an intriguing possibility of a relevance of quantum groups
in the T-duality and other superstring applications for example in
$D$-branes ~\cite{KlSWZ2}, ~\cite{KlSD}.

\vskip 10pt
\centerline{\bf ACKNOWLEDGEMENTS}
\frenchspacing
 I'm very gratefull to O. Andreev, B. Feigin, M. Lashkevich,
I. Polubin and V. Postnikov
for discussions. I would like to thank the Volkswagen-Stiftung
for financial support as well as
the members of the group of Prof. Dr. R. Shrader for
hospitality at Freie University of Berlin, where this work
was finished.
This work was supported in part by grant
INTAS-95-IN-RU-690.

\vfill
\end{document}